\documentclass[a4paper,fleqn,usenatbib]{mnras}

\usepackage{newtxtext,newtxmath}

\usepackage[T1]{fontenc}
\usepackage{ae,aecompl}

\usepackage{graphicx}	
\usepackage{amsmath}	
\usepackage{amssymb}	
\usepackage{MnSymbol}

\title[Metallicity inversion in the solar envelope]{Determining the metallicity of the solar envelope using seismic inversion techniques}

\author[G. Buldgen et al.]{
Ga\"el Buldgen$^{1}$\thanks{E-mail: gbuldgen@ulg.ac.be},
 S. J. A. J. Salmon$^{1}$,
 A. Noels$^{1}$,
 R. Scuflaire$^{1}$,
 M. A. Dupret$^{1}$, \and and
 D. R. Reese$^{2}$
\\
$^{1}$STAR Institute, Universit\'e de Li\`ege, All\'ee du Six Ao\^ut 19C, B-4000 Li\`ege, Belgium\\
$^{2}$LESIA, Observatoire de Paris, PSL Research University, CNRS, Sorbonne Universit\'e, UPMC Univ. Paris 06, Univ. Paris Diderot,\\ Sorbonne Paris Cit\'e, 5 place Jules Janssen, 92195 Meudon Cedex, France
}

\date{Accepted XXX. Received YYY; in original form ZZZ}

\pubyear{2017}

\begin{document}
\label{firstpage}
\pagerange{\pageref{firstpage}--\pageref{lastpage}}
\maketitle

\begin{abstract}
The solar metallicity issue is a long-lasting problem of astrophysics, impacting multiple fields and still subject to debate and uncertainties. While spectroscopy has mostly been used to determine the solar heavy elements abundance, helioseismologists attempted providing a seismic determination of the metallicity in the solar convective enveloppe. However, the puzzle remains since two independent groups prodived two radically different values for this crucial astrophysical parameter. We aim at providing an independent seismic measurement of the solar metallicity in the convective enveloppe. Our main goal is to help provide new information to break the current stalemate amongst seismic determinations of the solar heavy element abundance. We start by presenting the kernels, the inversion technique and the target function of the inversion we have developed. We then test our approach in multiple hare-and-hounds exercises to assess its reliability and accuracy. We then apply our technique to solar data using calibrated solar models and determine an interval of seismic measurements for the solar metallicity. We show that our inversion can indeed be used to estimate the solar metallicity thanks to our hare-and-hounds exercises. However, we also show that further dependencies in the physical ingredients of solar models lead to a low accuracy. Nevertheless, using various physical ingredients for our solar models, we determine metallicity values between $0.008$ and $0.014$.
\end{abstract}

\begin{keywords}
Sun: abundances -- Sun: fundamental parameters -- Sun: helioseismology -- Sun: interior -- Sun: oscillations
\end{keywords}



\section{Introduction} \label{sec:intro}
Helioseismology, the study of solar pulsations, has led to a number of successes. Amongst these achievements, one finds the determination of the position of the base of the solar convective envelope \citep[see][, and references therein.]{BasuConv}, the determination of the helium abundance in this region \citep{BasuY} as well as the demonstration of the necessity of microscopic diffusion to accurately reproduce the acoustic structure of the Sun \citep{BasuDiff}. Another important result of helioseismology was the determination of the solar rotation profile \citep{Brown, SchouRot,Schou}. However, the agreement that existed between solar models and structural inversions has been reduced since the publication of updated abundance tables by \citet{AspG} which showed a significant decrease of the metallicity with respect to the value commonly used in the standard solar models in the $90$s \citep[][ hereafter GN93 and GS98, respectively]{GN93, GreSauv}. These tables were further improved in \citet{AGSS} and have been at the center of what is now called the ``solar metallicity problem''. In $2011$, \citet{Caffau} published new abundance tables for which the metallicity was again re-increased. Attempts were also made to carry out seismic determinations of the solar metallicity, the first of which being that of  \citet{Takata} that hinted at a possible metallicity decrease $3$ years before the release of the new spectroscopic abundances, but could not conclude to the uncertainties in their inversion results. \citet{Basu} determined its value as $0.0172 \pm 0.002$ using an analysis of the dimensionless sound-speed gradient, denoted $W(r)$. In contrast, \citet{Vorontsov} determined that it should lie within $0.008$ and $0.013$ using a detailed analysis of the adiabatic exponent of solar envelope models for various equations of state.

In this study, we present a new method to derive an estimate of the metallicity in the solar envelope using the SOLA inversion technique \citep{Pijpers} and  the classical linear integral relations between frequency differences and structural corrections developped for metallicity kernels. In Sect. \ref{sec:method}, we present the structural kernels used and the target function of our metallicity estimate. We also present the possible causes of uncertainties in this method. In Sect. \ref{sec:harehounds}, we test our methodology in hare-and-hounds exercises and quantify the various contributions to the inversion errors. In Sect. \ref{sec:suninv}, we compute inversions of the solar metallicity for various trade-off parameters and reference models using different equations of state, we then discuss the reliability of our results and how further tests could complement and improve our study.
\section{Methodology} \label{sec:method}
To carry out our structural inversions, we use the linear integral relations obtained from the variational developments of the pulsation equations. The classical formulation of these relations is
\begin{align}
\frac{\delta \nu^{n,\ell}}{\nu^{n,\ell}}=\int_{0}^{1}K_{\rho,c^{2}}^{n,\ell}\frac{\delta \rho}{\rho}dx + \int_{0}^{1}K_{c^{2},\rho}^{n,\ell}\frac{\delta c^{2}}{c^{2}}dx,
\end{align}
which is used to carry out inversions of the solar sound speed profile. In this expression the notation $\delta$ denotes difference according to the following convention
\begin{align}
\frac{\delta x}{x} = \frac{x_{obs}-x_{ref}}{x_{ref}},
\end{align}
where $x$ can be an individual frequency of harmonic degree $\ell$ and radial order $n$ or a structural variable such as the density, $\rho$, or the squared adiabatic sound speed, $c^{2}=\frac{P}{\Gamma_{1}\rho}$, with $P$ the pressure and $\Gamma_{1}=\left( \frac{d \ln P}{d  \ln \rho}\right)_{S}$, the adiabatic exponent, where S is the entropy. The differences between structural variables are taken at fixed fractional radius $r/R$ with $R$ the total radius and $r$ the radial position. The $K^{n,l}$ functions are the structural kernels associated with the thermodynamical variables found in the variational expression. These functions only depend on equilibrium quantities and the eigenfunctions of the so-called reference model of the inversion.

\begin{figure*}
	\centering
		\includegraphics[width=14.5cm]{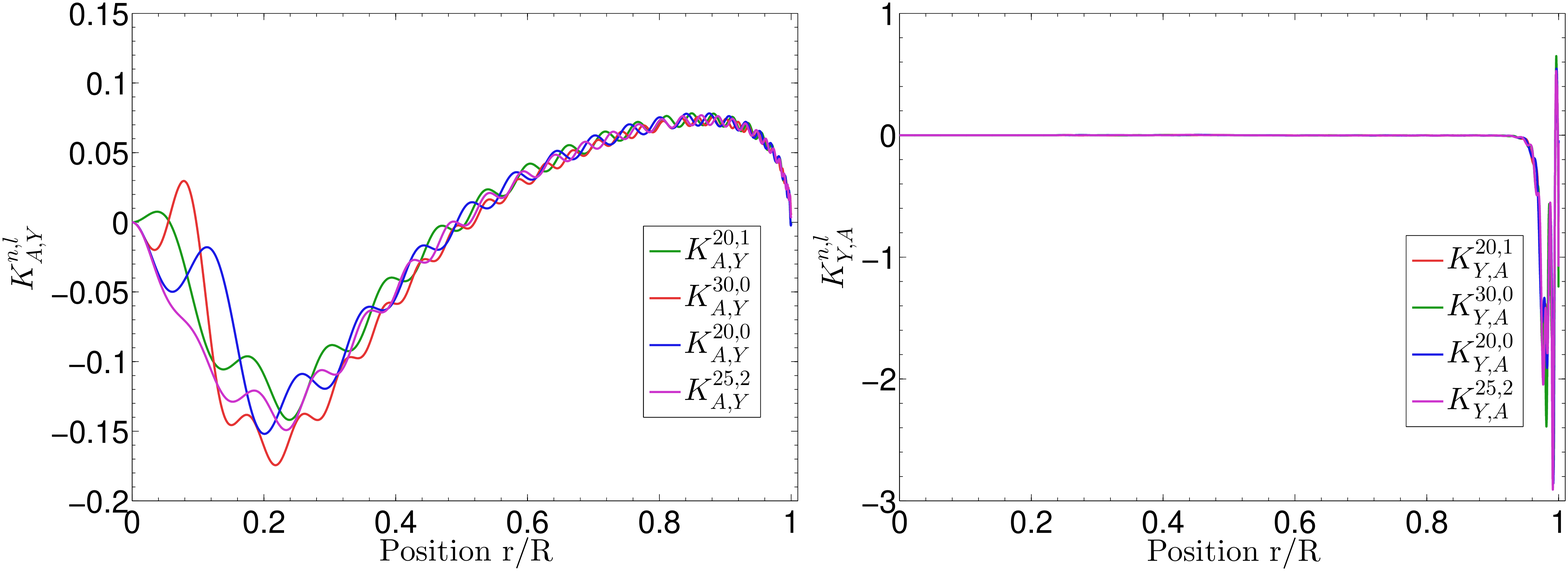}
		\includegraphics[width=14.5cm]{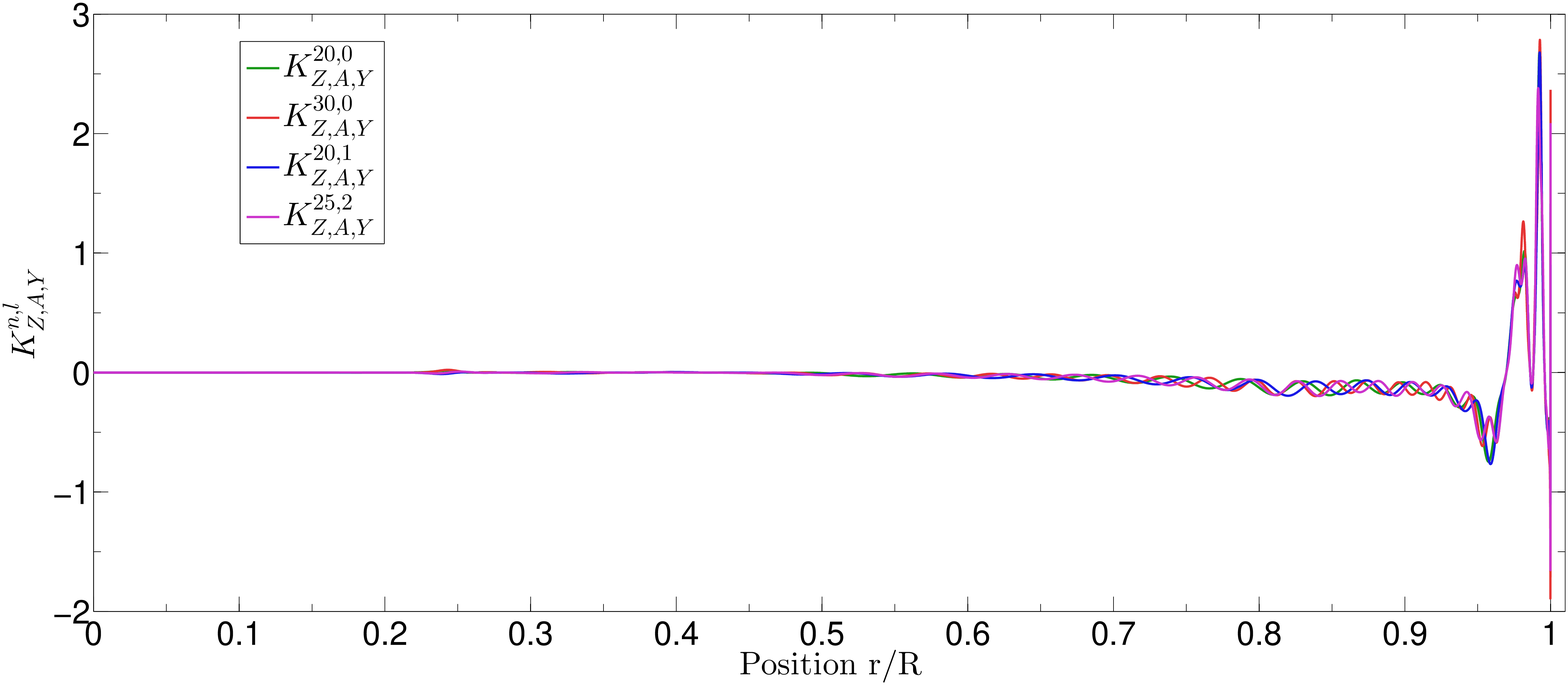}
	\caption{Kernels of the $(A,Y,Z)$ triplet for various oscillation modes (Upper left panel, $A$ Kernels) (Upper right panel, $Y$ kernels) (Lower panel, $Z$ kernels)}
		\label{figKerZ}
\end{figure*}

Using appropriate techniques \citep[see][]{BuldgenKer}, one can derive structural kernels for a large number of variables. Amongst them, one finds the kernels related to the convective parameter, $A$, defined as
\begin{align}
A=\frac{d \ln \rho}{d \ln r}-\frac{1}{\Gamma_{1}}\frac{d \ln P}{d \ln r},
\end{align}
presented in \citet{Elliott} as part of the $(A,\Gamma_{1})$ structural pair. It was initially used to test the equation of state. In this paper, we use the kernels of the $(A,Y,Z)$ triplet presented in \citet{BuldgenKer}, where $Y$ is the helium mass fraction and $Z$ the metallicity. We illustrate these kernels in figure \ref{figKerZ}. The first interesting point to notice is that the kernels associated with $A$ have a very small amplitude compared to those of $Y$ and $Z$. In addition, $A$ has the convenient property of being $0$ in the adiabatic region of the convective envelope. Similarly to \cite{Elliott} who considered the $(A,\Gamma_{1})$ kernels for carrying inversions of the $\Gamma_{1}$ profile in the Sun, the kernels we suggest in the present study can be used to efficiently carry out seismic inversions of both $Y$ and $Z$ in convective regions. The variational relation we use to carry out our inversion hence reads as
\begin{align}
\frac{\delta \nu^{n,\ell}}{\nu^{n,\ell}}=\int_{0}^{1}K_{A,Y,Z}^{n,\ell}\delta A dx + \int_{0}^{1}K_{Y,A,Z}^{n,\ell}\delta Y dx+ \int_{0}^{1}K_{Z,Y,A}^{n,\ell}\delta Z dx, \label{eqInvZ}
\end{align}
with $x$ the fractional radius.

\begin{figure*}
	\centering
		\includegraphics[width=13.5cm]{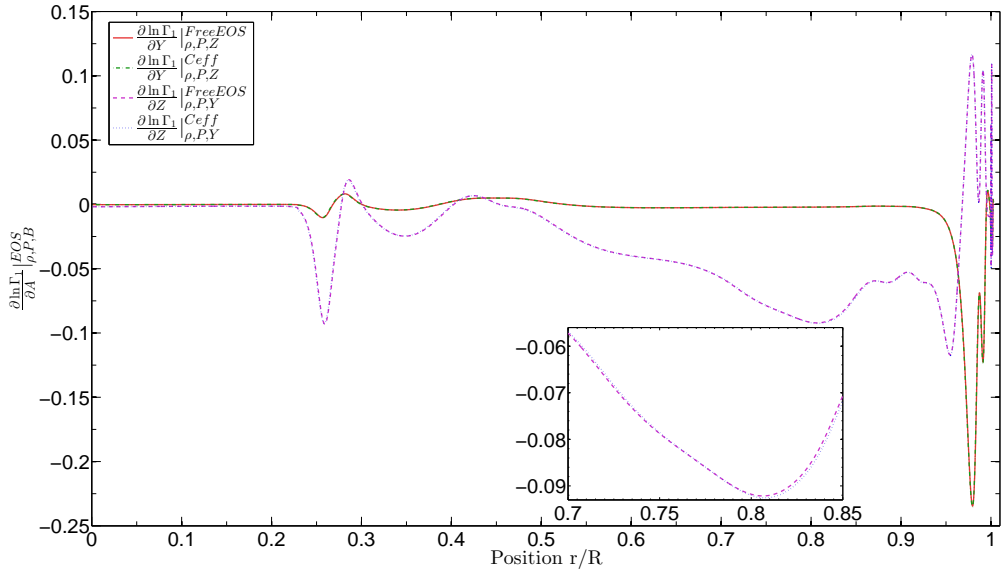}
	\caption{(Green and red curves) State derivatives of the natural logarithm of $\Gamma_{1}$ with respect to $Y$ for the FreeEOS and CEFF equations of state. (Magenta and blue curves) State derivatives of the natural logarithm of $\Gamma_{1}$ with respect to $Z$ for the FreeEOS and CEFF equations of state.}
		\label{figDGamm}
\end{figure*}

Although this study focusses on determinations of $Z$, we also carried out inversions of the helium abundance and found them to be in agreement with the classical helioseismic result of $0.2485\pm0.0035$, yet, less precise. As a comparison, \citet{Vorontsov} found that the helium mass fraction in the solar envelope was betweeen $0.24$ and $0.255$, which is also in agreement with the interval of $\left[0.242 \: 0.255 \right]$ we find. The main uncertainty in our determination is related to the physical ingredients in the models, leading to differences in $\Gamma_{1}$. Indeed, one uses the relation
\begin{align}
\frac{\delta \Gamma_{1}}{\Gamma_{1}}=&\left(\frac{\partial \ln \Gamma_{1}}{\partial \ln P}\right)_{\rho,Y,Z}\frac{\delta P}{P} + \left(\frac{\partial \ln \Gamma_{1}}{\partial \ln \rho}\right)_{P,Y,Z}\frac{\delta \rho}{\rho} + \left(\frac{\partial \ln \Gamma_{1}}{\partial Y}\right)_{P,\rho,Z} \delta Y \nonumber \\ 
&+ \left(\frac{\partial \ln \Gamma_{1}}{\partial Z}\right)_{P,\rho,Y} \delta Z \label{eq:Gamma1Eos}
\end{align}
to obtain perturbations of $Y$ and $Z$ in the variational expression. The main weakness in this approach is that the state derivatives of $\Gamma_{1}$ can vary not only with the equation of state but also with the calibration. Indeed, the calibration process will lead to slight differences in the acoustic structure when various physical ingredients (such as the opacity tables or the heavy elements abundance) are used. To assess this dependency, we use various physical ingredients in our hare-and-hounds exercises and our solar reference models used to determine the solar metallicity. A first illustration can be made, by comparing these derivatives for both the CEFF \citep{CEFF} and FreeEOS\footnote{http://freeeos.sourceforge.net/}\citep{Irwin} equations of state in a standard solar model. Both derivatives with respect to Y and Z are plotted in figure \ref{figDGamm} and illustrate that the general behaviour of both curves is extremely similar. However, differences of nearly $0.001$ can be seen in some regions and this of course means that, at the level of precision demanded, the choice of the equation of state will have an impact on the result. This is illustrated by the subplot of figure \ref{figDGamm}, where we take a closer look at the differences between the state derivatives with respect to Z.

We define the target function of our indicator as follows,
\begin{align}
\delta \mathcal{Z}= \frac{\int_{0}^{1} (1-x)^{2} \exp^{-800(x-0.79)^{2}} \delta Z dx}{\int_{0}^{1} (1-x)^{2} \exp^{-800(x-0.79)^{2}}dx} = \int_{0}^{1}\mathcal{T}_{\mathcal{Z}} \delta Z dx \label{eq:TarZ},
\end{align}
which of course implies $\delta \mathcal{Z} = \delta Z$ in the envelope up to an excellent accuracy, firstly because the chemical abundances are constant in the fully mixed convective zone of the Sun and secondly, the weight function of the indicator is virtually $0$ outside of a small interval (in normalised radius) within the envelope. The location of the Gaussian has been chosen such that it is not close to the helium ionization zone, which would imply a very strong correlation with the helium cross-term. The Gaussian width is also small enough so that there is no intensity in the radiative zone, which is not fully mixed. 

Looking at Figs. \ref{figKerZ} and \ref{figDGamm}, one can also see that this region is characterised by a rather large intensity in the $\Gamma_{1}$ derivatives with respect to $Z$, hence its associated kernels still having a non-negligible intensity around $0.8$ while the helium kernels are nearly $0$ at those depths. 

For this indicator, the cost function of the SOLA inversion is written as,
\begin{align}
\mathcal{J}_{\mathcal{Z}} = & \int_{0}^{1}\left[ K_{\mathrm{Avg}}-\mathcal{T}_{\mathcal{Z}}\right]^{2}dx +\beta \int_{0}^{1}K^ {2}_{\mathrm{Cross},Y}dx +\beta_{2} \int_{0}^{1}K^ {2}_{\mathrm{Cross},A}dx\nonumber \\
& + \tan(\theta) \sum^{N}_{i}(c_{i}\sigma_{i})^{2} +\lambda\left[1-\int_{0}^{1}K_{\mathrm{Avg}}dx\right], \label{eqCostSola}
\end{align}
where $K_{\mathrm{Avg}}$, $K_{\mathrm{Cross},A}$ and $K_{\mathrm{Cross},Y}$ are the so-called averaging and cross-term kernels \citep{Pijpers}. The averaging kernel is responsible for the matching of the target function of the inversion (thus the accuracy of the method). Meanwhile, the cross-term kernels are contributions for the additional variables found in Eq. \ref{eqInvZ} and which cannot be completely annihilated. For this particular inversion, two cross-terms have to be taken into account since we work with a triplet of variables instead of a pair. This means that the trade-off problem \citep{Backus} has to be analysed in depth and particular care has to be taken to check for possible compensation of the various error contributions in certain regions of the parameter space. The trade-off problem is assessed in the classical way, using the parameter $\theta$ to avoid a large contribution from the observational errors and the parameters $\beta$ and $\beta_{2}$ to reduce each cross-term contribution. The variable $\lambda$ is no free parameter, but a Lagrange multiplier associated with the unimodularity constraint.

The trade-off problem, and thus the quality of the inversion, is analysed by looking separately at the amplitudes of the terms associated with the averaging and cross-term kernels in Eq. \ref{eqCostSola}. These terms are denoted
\begin{align}
\vert \vert K_{avg} \vert \vert^{2}= & \int_{0}^{1}\left[ K_{\mathrm{Avg}}-\mathcal{T}_{\mathcal{Z}}\right]^{2}dx, \label{EqNormAVG}\\
\vert \vert K_{Cross,Y}\vert \vert^{2} = & \int_{0}^{1}K^{2}_{\mathrm{Cross},Y}dx, \label{EqNormCrossY}\\
\vert \vert K_{Cross,A}\vert \vert^{2}= & \int_{0}^{1}K^{2}_{\mathrm{Cross},A}dx. \label{EqNormCrossA}
\end{align}
However, these contributions do not contain all the information about the inversion technique. It is also useful to analyse separately each term of the inverted correction in hare-and-hounds exercises, to detect potential compensation effects. Therefore, we also analyse each error source of the inversion separately, using the definitions of \citet{Buldgentu}, which for this specific case are
\begin{align}
\epsilon_{\mathrm{AVG}}&=\int_{0}^{1}\left[ K_{\mathrm{Avg}}-\mathcal{T}_{\mathcal{Z}} \right] \delta Z dx, \label{EqEpsAVG} \\
\epsilon_{\mathrm{Cross},Y}&=\int_{0}^{1} K_{\mathrm{Cross,Y}} \delta Y dx, \label{EqEpsCrossY}\\
\epsilon_{\mathrm{Cross},A}&=\int_{0}^{1} K_{\mathrm{Cross,A}} \delta A dx, \label{EqEpsCrossA}\\
\epsilon_{\mathrm{Res}}&=Z_{Tar}-Z_{Ref} -\delta Z_{Inv}- \epsilon_{\mathrm{AVG}} - \epsilon_{\mathrm{Cross},Y} - \epsilon_{\mathrm{Cross},A} \label{EqEpsRes},
\end{align}
with $Z_{Tar}$ the metallicity of the target, which is known in a hare-and-hounds exercises, $Z_{Ref}$ the metallicity of the hound, $\delta Z_{Inv}$ the inverted correction found using the inversion. The residual error, $\epsilon_{\mathrm{Res}}$, which is obtained once all the other errors have been subtracted from the inversion results, encompasses all aspects that are not properly treated by the inversion. These include, in the hare-and-hounds exercises, the potential inadequacy of the surface effect corrections or the non-linear aspects stemming from the non-verification of the linear equations \ref{eqInvZ} or the assumptions on the equation of state made through Eq. \ref{eq:Gamma1Eos}.

\section{Hare-and-hounds exercises} \label{sec:harehounds}
Before carrying out the inversion on the actual solar seismic data, we carried out hare-and-hounds exercises to test the reliability of the metallicity determination. The stellar models used in the exercises were computed using the Li\`ege stellar evolution code (CLES, \citet{ScuflaireCles}). Their frequencies and eigenfunctions were computed using the Li\`ege adiabatic oscillation code (LOSC, \citet{ScuflaireLosc}). The inversions were computed using a customised version of the InversionKit software \citep{Reese}.

\subsection{Methodology and results}\label{Res}

We used a calibrated standard solar model built with the past solar GN$93$ \citep{GN93} abundances (high metallicity, Z) and the latest version of the Opal equation of state \citep{Rogerseos} as our hare. Furthermore, we used the latest OPLIB opacities \citep{Colgan} supplemented at low temperature by the opacities of \citet{Ferguson} and the effects of conductivity from \citet{Potekhin} and \citet{Cassisi}. The nuclear reaction rates implemented were those from the NACRE project \citep{Nacre}, supplemented by the updated reaction rate from \citet{Formicola} and we used the classical, local mixing-length theory \citep{Bohm} to describe convective motions. The hounds were fitted to the hare to ensure that they had exactly the same radii and similar luminosities. They were computed with the AGSS$09$ abundances and either the CEFF or the FreeEOS equation of state. There are two reasons justifying the fact that we did not use the Opal equation of state in our reference models. Firstly, this allows us to test for the biases stemming from the equation of state in our inversions. Secondly, the Opal equation of state, being defined on a grid with finite resolution, is not well suited to compute the $\Gamma_{1}$ derivatives numerically with high accuracy. The CEFF and FreeEOS equations of state, being defined with analytical relations, do not suffer from this problem and were thus favoured. 

We summarise the properties of the hare and the hounds in table \ref{tabHoundsProp}. As can be seen, some of these reference models differ significantly from the target. These differences result from the way the hounds were generated, which was not a classical solar calibration as is usually done for the Sun. All hounds reproduce accurately the radius of target $1$, a given value of helium abundance in the convective envelope and the position of its base within a less constraining accuracy. We did not limit the values of helium abundance and position of the base of the convective envelope to those of target $1$ to assess the effects of those constraints on the cross-term contributions. To further assess the uncertainties, additional changes were induced in Ref $1$, $2$ and $3$ by not including microscopic diffusion of the heavy elements. The changes in the position of the base of the convective envelope for the various hounds were induced by changing the value of an undershoot parameter within values up to $0.2$ of the local pressure scale height before recalibrating the hounds. As can be seen from the luminosity and age values in table \ref{tabHoundsProp}, this calibration method can lead to quite large differences between the models, hence sometimes implying larger differences in acoustic structure between the hare and the hounds than in a classical solar calibration approach. 

We used the radius, the helium abundance in the convective envelope and the position of its base as constraints for the fit of the hounds to the hare. Once a satisfying fit was obtained, we changed an undershoot parameter and the diffusion velocities, while still fitting the radius, to be able to analyze the behaviour of the inversion when facing changes in $A$ and $Y$. Consequently, some non-linearities could be observed for a few of these perturbed models and could help us understand the cases were the inversion was not so stable. To be as close to reality as possible, we used the exact same set of low $\ell$ oscillation modes observed for the Sun as presented in \citet{BasuFreq}, namely $2189$ oscillation modes with $\ell$ from $0$ up to $250$, and used the same uncertainties on the frequencies as those of the actual solar modes. We tested a total of $17$ hounds and analysed whether the inversion was efficient in determining the metallicity of the hare. Due to the large differences between the hitherto seismically determined values of the solar metallicity, we wanted to assess whether our method would be able to distinguish between several solutions. 
\begin{table*}
\caption{Characteristics of the models used in the hare-and-hounds exercises}
\label{tabHoundsProp}
  \centering
\begin{tabular}{r | c | c | c | c | c | c | c | c}
\hline
& Mass $(M_{\odot})$  & Age $(Gy)$ & Radius $(R_{\odot})$ & $L$ $(L_{\odot})$ & $Y_{CZ}$ & $Z_{CZ}$ & $\left(\frac{r}{R}\right)_{CZ}$ & EOS\\ \hline
Target $1$ & $1.0$ & $4.5794$ & $1.0$ & $0.9997$ & $0.2397$ &$0.01818$&$0.07087$&Opal$05$\\
Ref 1 & $1.0$ & $4.8073$ & $1.0$ & $0.9564$ & $0.2307$ & $0.01513$&$0.07087$&FreeEOS\\  
Ref 2 & $1.0$&  $4.8038$ & $1.0$& $0.9558$& $0.2294$ &$0.01513$&$0.7125$
&FreeEOS\\  
Ref 3 & $1.0$& $4.5097$ & $1.0$& $0.9740$ & $0.2342$ &$0.01513$&$0.716$
&FreeEOS\\
Ref 4 & $1.0$ & $4.2502$ & $1.0$ & $0.9933$ & $0.2421$&$0.01513$&$0.7068$&FreeEOS\\
Ref 5 & $1.0$ & $4.4130$ & $1.0$ & $1.0107$ &$0.2455$& $0.01373$&$0.7161$
&FreeEOS\\
Ref 6 & $1.0$ & $5.0004$ &$1.0$ & $0.9749$ & $0.2301$& $0.01375$&$0.7087$&FreeEOS\\
Ref 7 & $1.0$ & $4.8743$ &$1.0$ & $0.9817$ & $0.2300$& $0.01373$&$0.7171$&FreeEOS\\
Ref 8 & $1.0$ & $4.4081$ &$1.0$ & $1.0100$ & $0.2419$& $0.01381$&$0.7066$&FreeEOS\\
Ref 9 & $1.0$ & $4.5285$ &$1.0$ & $1.0035$ & $0.2381$& $0.01369$&$0.7160$&FreeEOS\\
Ref 10 & $1.0$ & $5.3522$ &$1.0$ & $0.9695$ & $0.2285$& $0.01358$&$0.7090$&CEFF\\
Ref 11 & $1.0$ & $5.2909$ &$1.0$ & $0.9731$ & $0.2282$& $0.01349$&$0.7150$&CEFF\\
Ref 12 & $1.0$ & $5.2253$ &$1.0$ & $0.9761$ & $0.2283$& $0.01345$&$0.7170$&CEFF\\
Ref 13 & $1.0$ & $4.7455$ &$1.0$ & $1.0046$ & $0.2402$& $0.01373$&$0.7055$&CEFF\\
Ref 14 & $1.0$ & $4.8635$ &$1.0$ & $0.9982$ & $0.2364$& $0.01361$&$0.7160$&FreeEOS\\
Ref 15 & $1.0$ & $4.7433$ &$1.0$ & $1.0054$ & $0.2283$& $0.01355$&$0.7236$&FreeEOS\\
Ref 16 & $1.0$ & $4.5651$ &$1.0$ & $1.0155$ & $0.2435$& $0.01376$&$0.7075$&FreeEOS\\
Ref 17 & $1.0$ & $4.5790$ &$1.0$ & $0.9993$ & $0.2435$& $0.01344$&$0.7120$&FreeEOS\\
\hline
Target $2$ & $1.0$ & $4.5787$ & $1.0$ & $1.0002$& $0.2386$ & $0.01820$  & $0.7080$ & FreeEOS \\
\hline
\end{tabular}
\end{table*}

\begin{figure*}
	\centering
		\includegraphics[width=13.5cm]{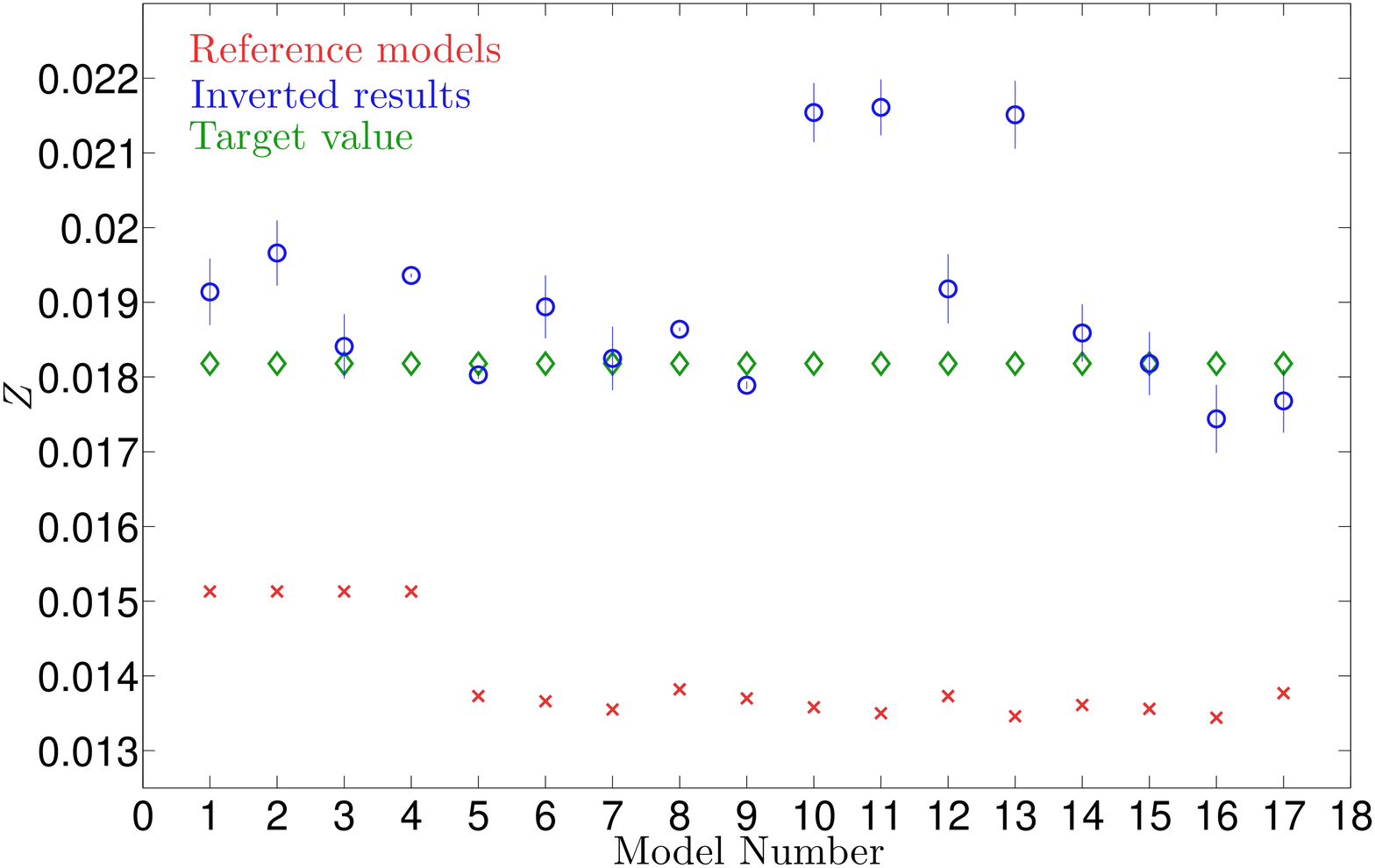}
	\caption{Inversion results for the hare-and-hounds exercises between the $17$ reference models and the target model of table \ref{tabHoundsProp}.}
		\label{figResHH}
\end{figure*}

The results of these inversions are given in table \ref{tabResInvHH} and illustrated in figure \ref{figResHH}. For all the cases illustrated here, the inversion could determine an estimate of the metallicity in the convective envelope. However, one can see that for reference models (hounds) $2$, $10$, $11$ and $13$, these estimates are not so accurate. Globally, a spread of around $4 \times 10^{-3}$ is found for the inversion results. This spread is due to multiple sources: e.g., for models $10$, $11$ and $13$, the equation of state was the CEFF equation of state, which leads to much larger differences in $\Gamma_{1}$ with the Opal$05$ equation of state than does the FreeEOS equation of state. Besides the differences from the equation of state, intrinsic differences in $\Gamma_{1}$ due to differences in stratification of the model, can have a strong impact on the accuracy of the method. It is important to bear in mind that the variables Y and Z are not directly seismically constrained but are obtained via an equation of state which is furthermore linearised using to Eq. \ref{eq:Gamma1Eos}. Consequently, errors in the derivatives of $\Gamma_{1}$ can quickly arise and reduce the accuracy of the results. Moreover, hounds having a poor representation of the stratification of the hare model will also have rather large $A$ differences and thus intrinsically larger cross-term contributions. In Sect. \ref{Stab}, we further investigates this point and discuss the determination of the trade-off parameters and what they can teach us about the characteristics of our approach.
\begin{table*}
\caption{Inversion results for the hare-and-hounds exercises.}
\label{tabResInvHH}
  \centering
\begin{tabular}{r | c | c | c | c | c}
\hline
& $Z_{Ref}$  & $Z_{Inv}$ & $\vert \vert K_{Avg}-\mathcal{T}_{\mathcal{Z}} \vert \vert^{2}$ & $\vert \vert K_{Cross,Y} \vert \vert^{2}$ & $\vert \vert K_{Cross,A} \vert \vert^{2}$ \\ \hline
Ref 1 & $0.01513$ & $0.01914\pm 4.35 \times 10^{-4}$ & $0.0211$ & $0.07683$ & $5.000\times 10^{-3}$\\  
Ref 2 & $0.01513$&  $0.01996\pm 4.26 \times 10^{-4}$ & $0.02198$& $0.07680$& $4.53\times 10^{-3}$\\  
Ref 3 & $0.01513$& $0.01841\pm 4.25 \times 10^{-4}$ & $0.02169$& $0.07602$ & $4.549 \times 10^{-3}$ \\
Ref 4 & $0.01513$ & $0.01936 \pm 2.15 \times 10^{-5}$ & $0.02230$ & $0.07538$ & $1.816$\\
Ref 5 & $0.01373$ & $0.01803 \pm 5.09 \times 10^{-5}$ & $0.1245$ & $0.07144$ &$1.409$\\
Ref 6 & $0.01366$ & $0.01894 \pm 4.15 \times 10^{-4}$ &$0.0254$ & $0.07376$ & $0.02959$\\
Ref 7 & $0.01355$ & $0.01825\times 4.15\times 10^{-4}$ &$0.02036$ & $0.07321$ & $1.967\times 10^{-2}$\\
Ref 8 & $0.01382$ & $0.01864 \pm 2.14\times10^{-5}$ &$0.02026$ & $0.07253$ & $1.820$\\
Ref 9 & $0.01370$ & $0.01789 \pm 5.09\times 10^{-5}$ &$0.1245$ & $0.07145$ & $1.410$\\
Ref 10 & $0.01358$ & $0.02154\pm 3.87\times 10^{-4}$ &$0.1230$ & $0.07182$ & $0.03715$\\
Ref 11 & $0.01350$ & $0.02161 \pm 3.71 \times 10^{-4}$ &$0.05124$ & $0.07404$ & $0.04631$\\
Ref 12 & $0.01373$ & $0.01918\pm 4.57 \times 10^{-4}$ &$0.01603$ & $0.07280$ & $3.552\times 10^{-4}$\\
Ref 13 & $0.01346$ & $0.02151 \pm 4.49 \times 10^{-4}$ &$0.01666$ & $0.07348$ & $1.516\times 10^{-3}$\\
Ref 14 & $0.01361$ & $0.01859 \pm 3.76 \times 10^{-4}$ &$0.05211$ & $0.07313$ & $0.02439$\\
Ref 15 & $0.01356$ & $0.01818 \pm 4.23 \times 10^{-4}$ &$0.02022$ & $0.07238$ & $4.572\times 10^{-3}$\\
Ref 16 & $0.01344$ & $0.01744 \pm 4.48\times 10^{-4}$ &$0.01642$ & $0.07257$ & $1.509\times 10^{-3}$\\
Ref 17 & $0.01377$ & $0.01768\pm 4.24\times 10^{-4}$ &$0.0228$ & $0.07246$ & $4.578\times 10^{-3}$\\
\hline
\end{tabular}
\end{table*}

\subsection{Analysis of the stability of the inversion technique}\label{Stab}
\subsubsection{Quality of the kernel fits}
We have seen in Sect. \ref{Res} that an indication of the metallicity of the Sun can be provided by seismology. However, inversion techniques can sometimes provide accurate results because of compensation of their errors. Therefore, their reliability and stability should be properly assessed.  

\begin{figure*}
	\centering
		\includegraphics[width=13.5cm]{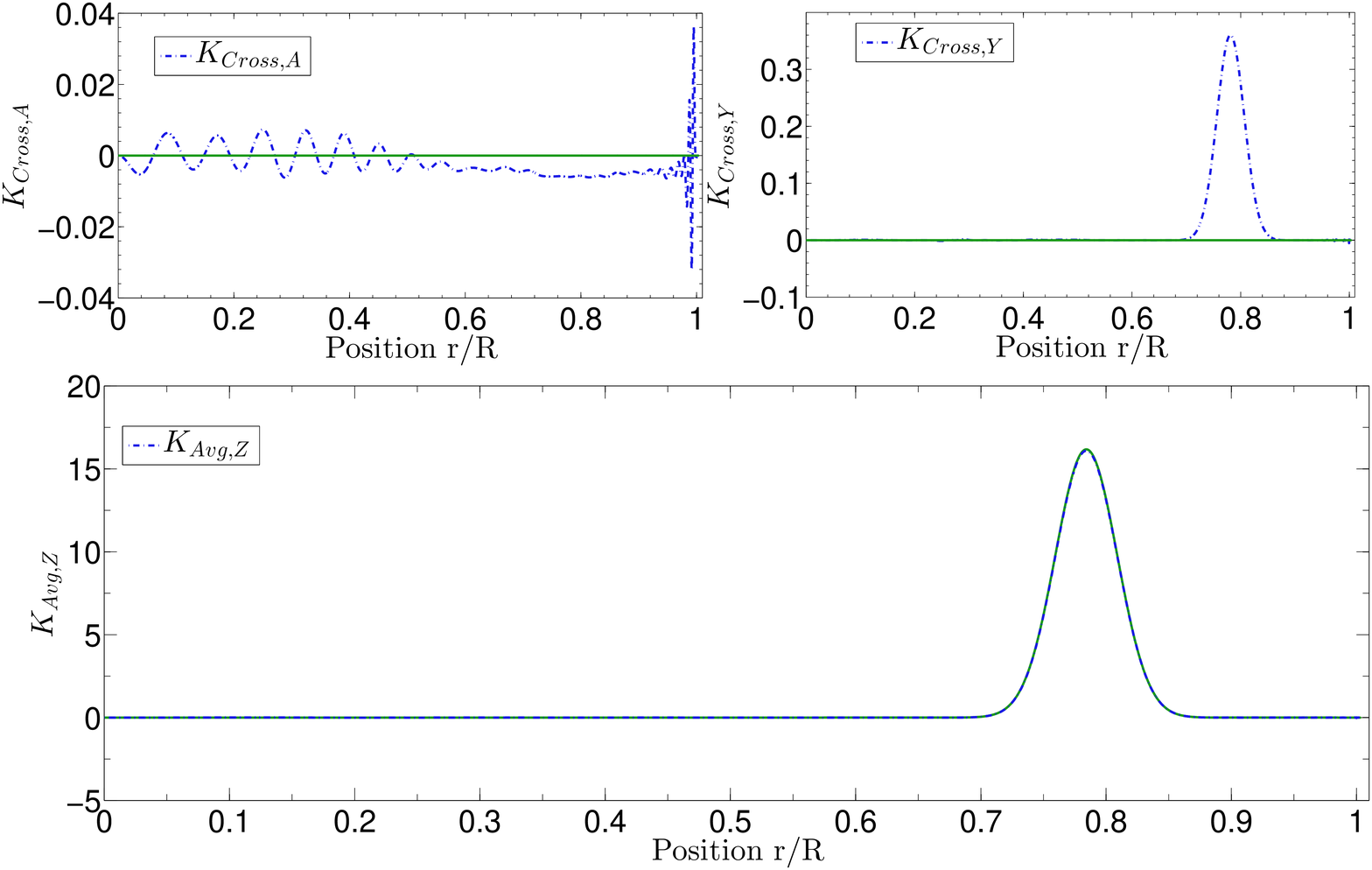}
	\caption{(Upper-left panel) Cross-term associated with $A$ in blue and the target function (being $0$ here) in green. (Upper-right panel) Cross-term associated with $Y$ in blue and the target function (being $0$) in green. (Lower panel) Averaging kernel of the inversion in blue and the respective target function in green.} 
		\label{figAvgCross}
\end{figure*}

\begin{figure*}
	\centering
		\includegraphics[width=13.5cm]{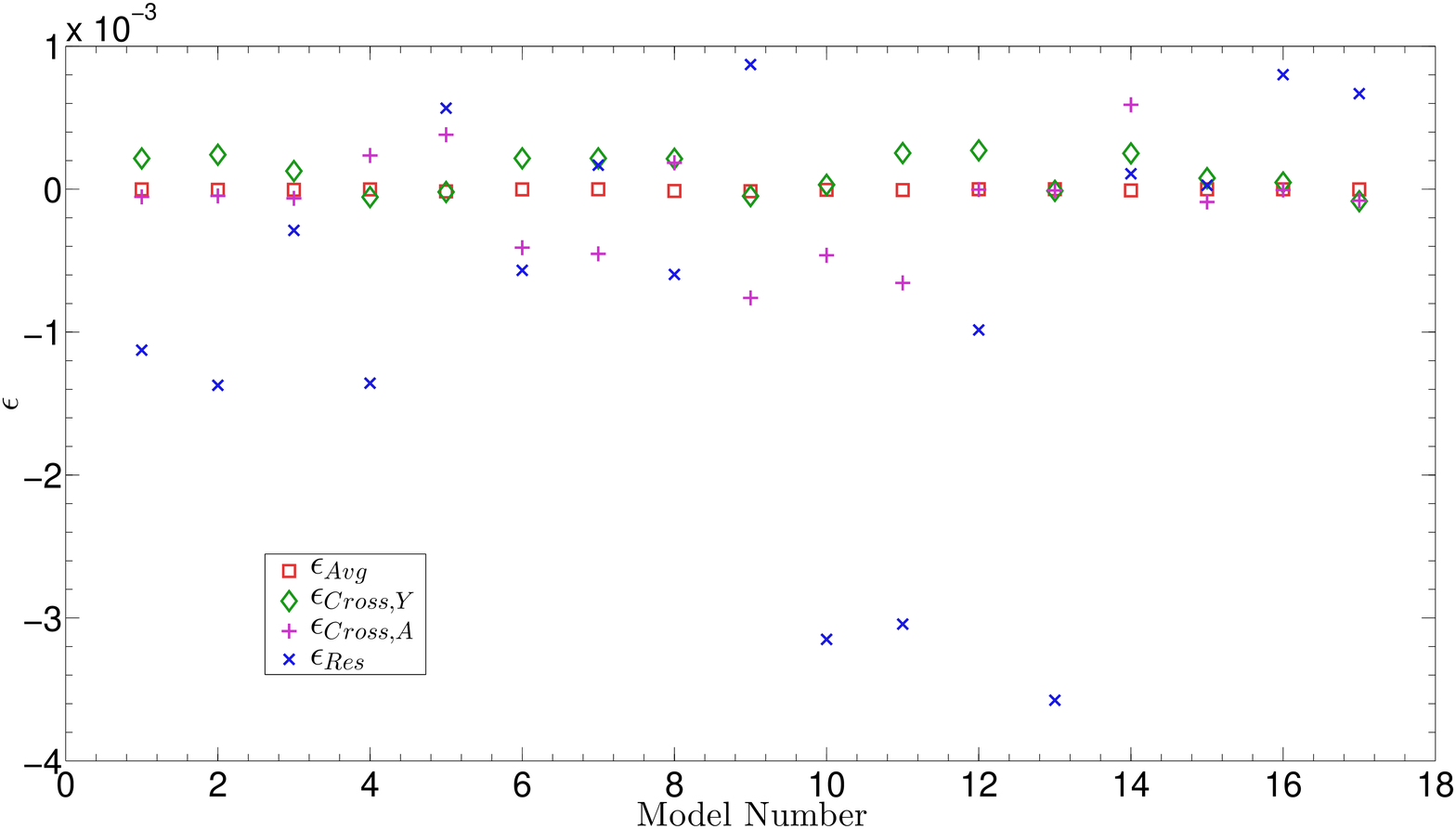}
	\caption{Error contributions, denoted $\epsilon_{Avg}$,$\epsilon_{CrossY}$,$\epsilon_{Cross,A}$,$\epsilon_{Res}$} for the hare-and-hounds exercises. 
		\label{figErrorContrib}
\end{figure*}

We propose to first assess the quality of the kernel fits by looking at the amplitude of the first three terms of Eq. \ref{eqCostSola}, in the last columns of table \ref{tabResInvHH}. To give a better idea of what these represent, we illustrate in figure \ref{figAvgCross} the fit of the averaging kernel and the cross-term kernels for the inversion of Ref $3$. From a first visual inspection, the fit of the averaging kernel seems good and the contribution from the cross-term associated with $A$ seems close to $0$. By inspecting the helium cross-term kernel, we can see a strong correlation with the averaging kernel. However, the height of this peak is around $45$ times smaller than the height of the averaging kernel. This means that the helium contribution should be kept small, as can be seen in figure \ref{figErrorContrib}. This difference in amplitude is a consequence of the choice of the position for the peak in the target function. Had we chosen a target much closer to the helium ionization zones, then the correlation with the helium kernels would have been much more problematic. 

To analyse whether compensation is present in the inversion, we plot in figure \ref{figErrorContrib} the real error contributions for all hounds. The kernel fits are informative of the quality of the inversion. However plotting the real error contributions can indicate whether what appears visually as a satisfactory fit is actually the reason for the accuracy of the result. 

\subsubsection{Analysis of the parameter space}
\begin{figure*}
	\flushleft
		\includegraphics[width=18.5cm]{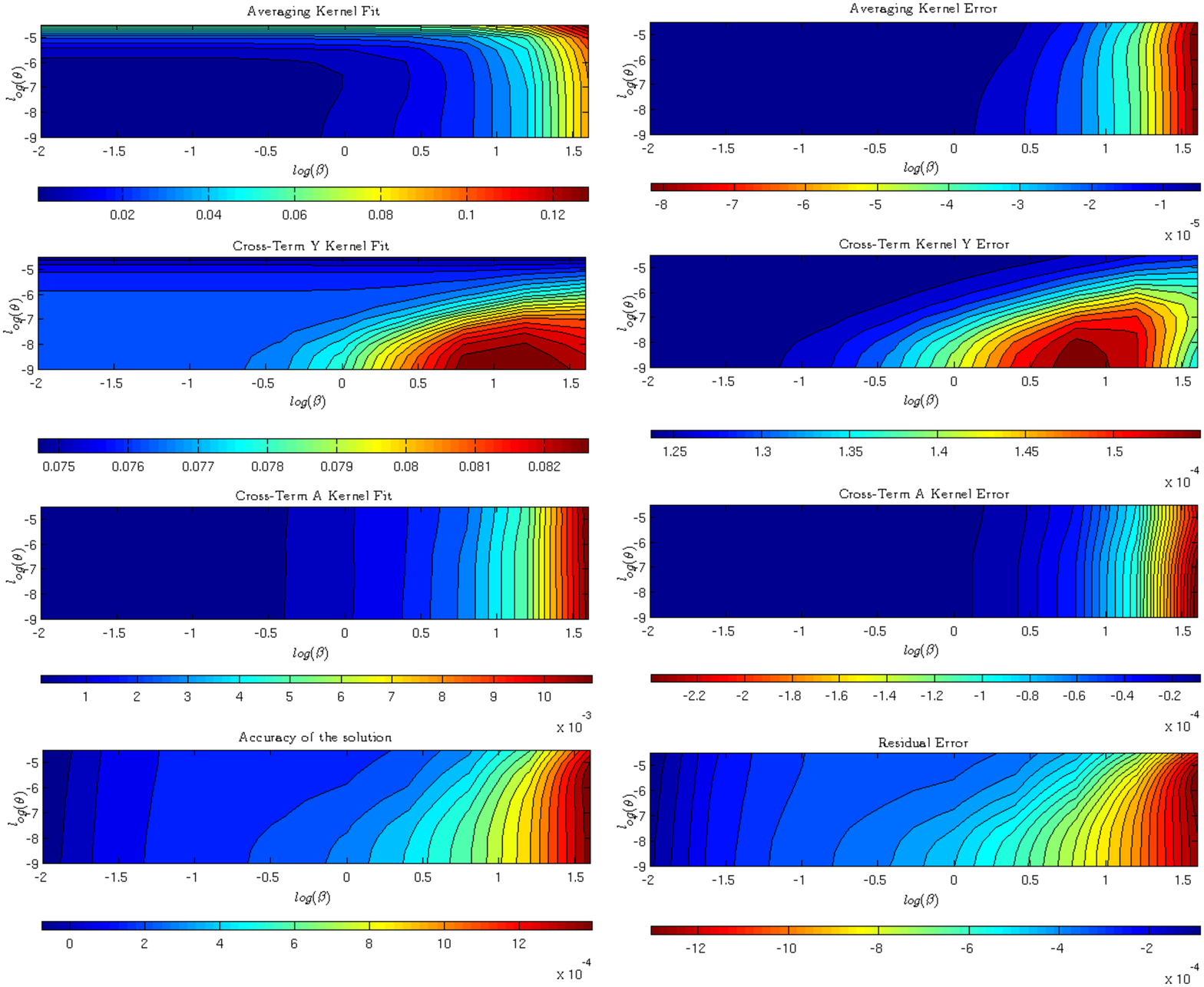}
		\centering
		\includegraphics[width=11.8cm]{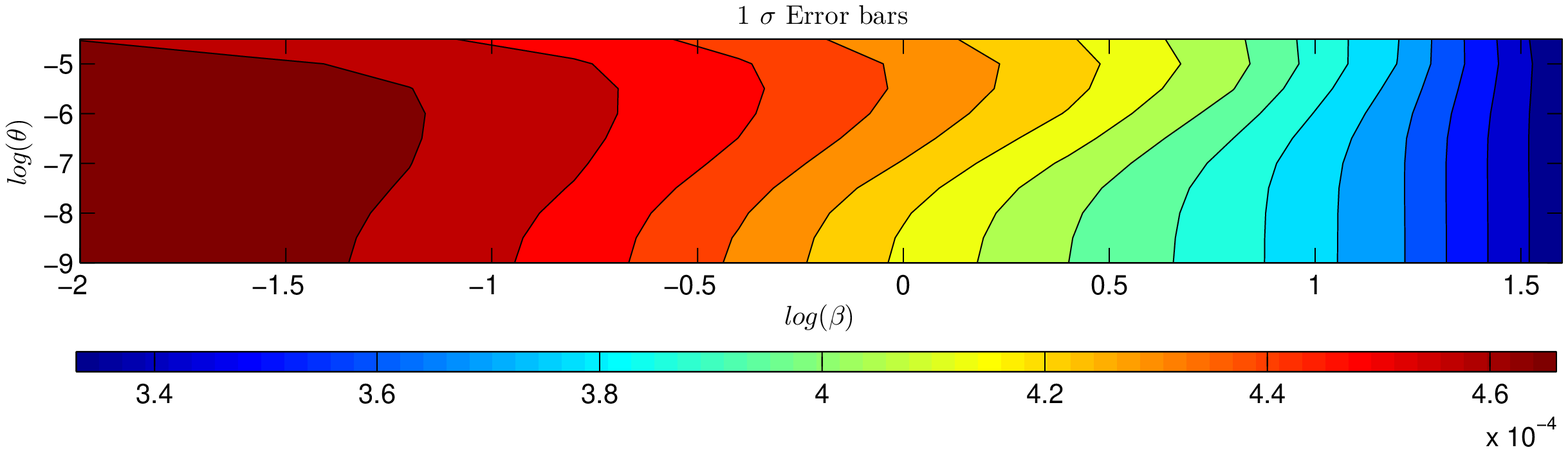}
	\caption{Variations of various quality checks during scan of the $\beta-\theta$ plane for a fixed $\beta_{2}$ value of $1$. (Upper line left) $\vert \vert K_{Avg}-\mathcal{T}_{\mathcal{Z}} \vert \vert^{2}$, defined in Eq. \ref{EqNormAVG},  (Upper line right) $\epsilon_{Avg}$, defined in Eq. \ref{EqEpsAVG}, (Second line left) $\vert \vert K_{Cross,Y} \vert \vert^{2}$, defined in Eq. \ref{EqNormCrossY}, (Second line right) $\epsilon_{Y}$, defined in Eq. \ref{EqEpsCrossY}, (Third line left) $\vert \vert K_{Cross,A} \vert \vert^{2}$, defined in Eq. \ref{EqNormCrossA}, (Third line right) $\epsilon_{A}$, defined in Eq. \ref{EqEpsCrossA}, (Fourth line left)  Accuracy of the result, defined as $Z_{Tar}-Z_{Inv}$, (Fourth line right) $\epsilon_{Res}$, defined in Eq. \ref{EqEpsRes}, (Lower plot) $1$ $\sigma$ error bars of the inversion results, illustrated here for Ref $3$ (The blue regions are associated with smaller errors.).}
		\label{figScan}
\end{figure*}

The main difficulty of carrying out the inversion is to find a given set of parameters for which the real errors of the inversion, here defined as the $\epsilon_{i}$ in Eqs. \ref{EqEpsAVG} to \ref{EqEpsRes}, are small and do not present compensation effects, and where the squared norm of the differences between the averaging and cross-term kernels, defined in Eqs. \ref{EqNormAVG} to \ref{EqNormCrossY}, and their respective targets have reduced amplitudes. Depending on the quality of the reference model or the dataset, the trade-off problem will be different and multiple sets of parameters can be found. 
Therefore, to further understand the reliability of the inversion and its parameter dependency, we did for Ref $3$ a brief scan of the parameter space, which is illustrated in figure \ref{figScan}. Firstly, we note from the lowest left panel in figure \ref{figScan}, illustrating the differences between the inverted $Z$ value and the target value of $Z$, that the changes in the parameter do not induce a large deviation of the inversion results from the value found with the optimal set of parameters. However, other inversions in our hare-and-hounds exercises were not as stable, mainly due to an increase of the residual error or to the cross-term contribution associated with the convective parameter. For this scan, we fixed a value of $\beta_{2}$ to $1.0$ and varied $\theta$ from $10^{-4}$ to $10^{-9}$ and $\beta$ from $10^{-2}$ to $10^{1.5}$ (see eq. \ref{eqCostSola}). A first observation that can be made is that the variations of all quantities are regular with the parameters. Sudden and steep variations in the accuracy of the solution and the real error contributions would indicate that the inversion is not sufficiently regularised. In such cases, one has no real control on the trade-off problem and thus no means of ensuring a correct result.

In figure \ref{figScan}, we can see that for a fixed value of $\beta_{2}$ the optimum is found for a low $\theta$ and a low $\beta$. However, the low $\beta$ value has to be considered carefully when the helium abundance differences are larger. For this particular test between Ref $3$ and Target $1$, the differences is of the order of $0.005$, thus intrinsically quite small, which naturally reduces the value of $\epsilon_{Cross, Y}$. In addition to the helium cross-term contribution, the amplitude of the error bars could be strongly increased by the reduction of the $\theta$ parameter. In this particular case, the lower plot of figure \ref{figScan} shows that the error bars remained quite small for the whole scan but this will not necessarily be the case for other inversions.

Overall, we can see from the lower plot of figure \ref{figScan} that the inversion is very stable, since the errors remain of the order of $10^{-4}$, with the exception of the residual error that can reach is this case the order of $10^{-3}$ for a particular region of the parameter set. This region is related to high values of $\beta$ and $\beta_{2}$ (which is fixed) and a range of low $\theta$ values of around $10^{-8}$ and $10^{-9}$.
This is a consequence of the regularising nature of the $\theta$ parameter, which is used to adjust the trade-off between precision and accuracy of the method. In this particular case, the instability of the inversion is a consequence of the very low weight given to the damping of the error bars, and thus the precision of the method. A similar behaviour can also be observed in solar inversions of full structural profiles, were the observational error bars remain small, but the profile already shows an unphysical oscillatory behaviour. 

\begin{flushleft}
\textbf{$\beta$ and $\beta_{2}$ Parameter}
\end{flushleft}
Our hare-and-hound exercises allowed us to define various parameter values for which the inversion was stable, depending on the reference model quality. In most cases, the value of $\beta$, the free parameter associated with the helium contribution, was set between approximately $0.1$ and $10$, as can be seen in table \ref{tabParamsInvHH}, depending on the intrinsic difference in helium content between the target and the reference model. Indeed, if the helium abundance value is naturally close to that of the target, it is unnecessary to strongly damp the helium cross-term and penalise the elimination of the convective parameter cross-term. Actually, it should be noted that $\beta$ and $\beta_{2}$ are correlated. As damping the helium cross-term is done by reducing the amplitude of the inversion coefficients, it can sometimes have a similar effect on the convective parameter cross-term.

\begin{figure*}
	\centering
		\includegraphics[width=13.5cm]{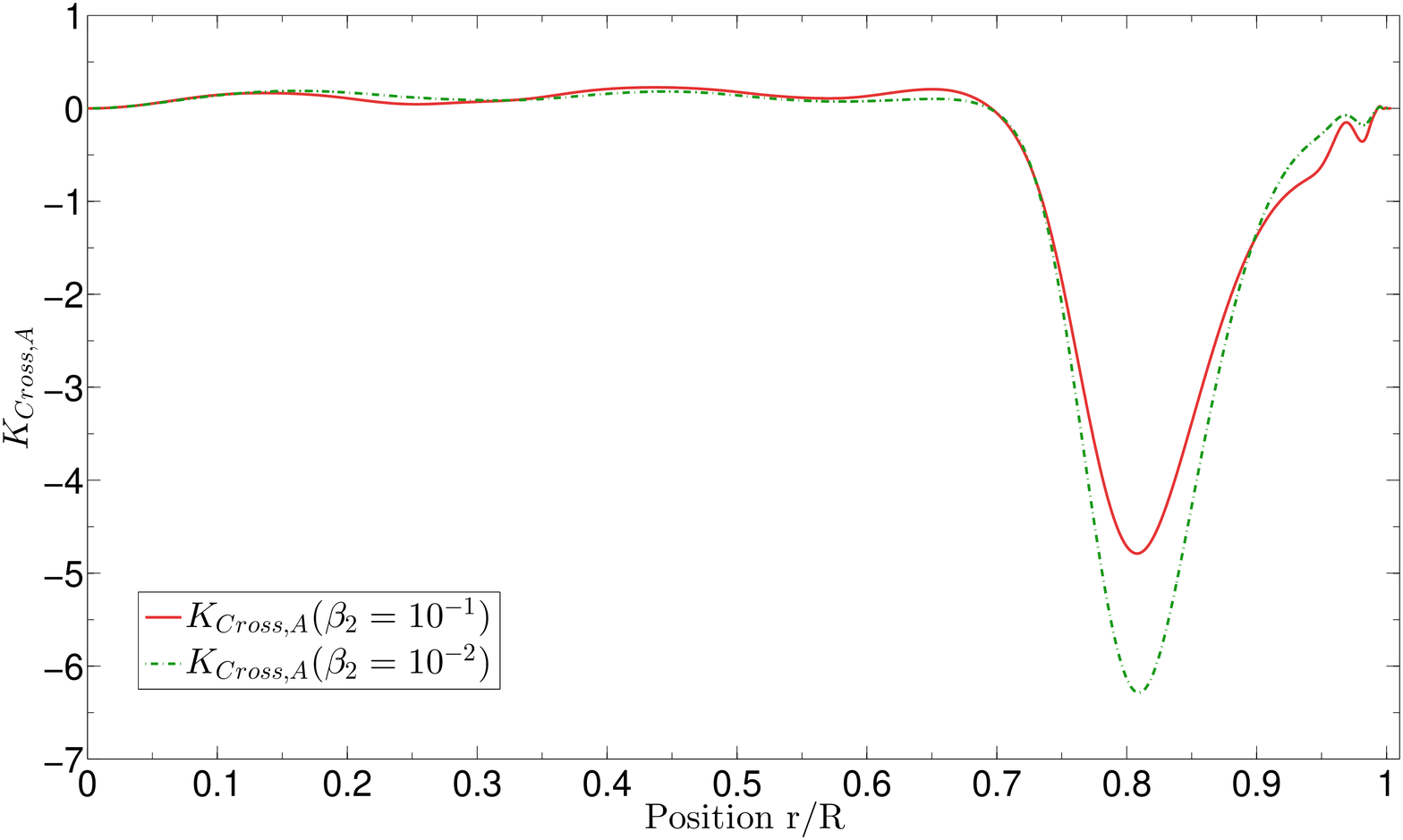}
	\caption{Cross-terms from inversions with a differente $\beta_{2}$ parameter. The red curve, associated with a high $\beta_{2}$ shows a lower amplitude than the green curve in the convective zone, but a higher amplitude in the radiative zone, where $A$ is non zero.}
		\label{figCrossA}
\end{figure*}

The value of $\beta_{2}$, associated with the cross-term contribution of $A$, was set to $0.1$ most of the time, although some inversions required a value as high as $1$ to efficiently damp the cross term contributions and others only required a value as low as $0.01$. The values of $\beta_{2}$ for each individual inversion illustrated in figure \ref{figResHH} are also illustrated in table \ref{tabParamsInvHH}. An important remark has to be made about the cross-term associated with $A$. Due to the fact that $A$ is exactly $0$ in the convective envelope, damping the cross-term in $A$ using $\beta_{2}$ can have unwanted consequences, as damping the total integral does not necessarily mean that the real contribution to the error, stemming from the radiative zone, will be always damped. For example, the inversions associated with a low $\beta_{2}$ parameter show a high peak in the convective envelope that contributes to the norm $\vert \vert K_{Cross,A} \vert \vert^{2}$, but does not contribute to the real error, $\epsilon_{Cross,A}$. Increasing $\beta_{2}$ reduces the peak in the convective zones but increases the amplitude of the cross-term kernel in the radiative zone which leads to an increase of the total error This effect is illustrated in figure \ref{figCrossA}. However, systematically keeping very low values of $\beta_{2}$ can also imply large contributions from the cross-term associated with $A$, and thus an inaccurate inversion. It is clear that some compensation in the integral definition of $\epsilon_{Cross,A}$ is responsible for the very low error contribution for the models having larger $\vert \vert K_{Cross,A} \vert \vert^{2}$, namely Refs $4$, $5$, $8$ and $9$. 

\begin{flushleft}
\textbf{$\theta$ Parameter}
\end{flushleft}

From table \ref{tabParamsInvHH}, we can see that optimal value of the $\theta$ parameter was found to vary around $10^{-6}$ for most inversions. Four cases showed higher optimal values of $\theta$, going up to $10^{-3}$, marking a clear separation with most of the parameter sets which are very similar. All these models showed slightly unstable behaviours and a higher $\theta$ value seems logical since it is a regularization parameter. One can also point out that Refs $4$ and $8$ had much larger individual frequency differences\footnote{Ranging from a factor $2$ up to a factor $10$ when compared to individual differences of models such as model $1$ or $7$.} with the hare. This implies that the differences in acoustic structure between these models and the hare are larger than between other references models and the hare. This is confirmed by the observation that these models showed larger discrepancies in $\left(\frac{\partial \ln \Gamma_{1}}{\partial Z}\right)_{P,\rho,Y}$ at the point where the target function is located. These discrepancies are the main reason for the inaccuracies of the method, since this factor directly multiplies the estimated metallicity. They can arise from differences in $\Gamma_{1}$ present either in the equation of state or stemming from the calibration. These differences can also be seen in sound speed and density profile comparisons. Physically, the reason why Ref $4$ or Ref $8$ shows larger discrepancies than Ref $1$ or Ref $6$ is linked to the temperature gradient in the radiative region of the models. Indeed, Ref $1$ (respectively Ref $6$) has the same metallicity than Ref $4$ (respectively Ref $8$) but has a lower helium abundance, and thus a higher hydrogen abundance meaning that overall, the opacity in the radiative region will be increased and thus lead to an acoustic structure in better agreement with that of Target $1$ which has a high metallicity.

\begin{table*}
\caption{Trade-off parameters values for the hare-and-hounds exercises.}
\label{tabParamsInvHH}
  \centering
\begin{tabular}{r | c | c | c }
\hline
& $\theta$  & $\beta$ & $\beta_{2}$  \\ \hline
Ref 1,2,3,6,7,12,15,17 & $10 ^{-6}$ & $10$ & $1.0$ \\  
Ref 4,8 & $10 ^{-3}$ & $10$ & $10^{-3}$ \\
Ref 5 & $10 ^{-3}$ & $10^{-1}$ & $10^{-1}$ \\
Ref 9 & $10 ^{-3}$ & $1$ & $0.1$ \\
Ref 10 & $10 ^{-5}$ & $10$ &$0.1$ \\
Ref 11 & $10 ^{-5}$ & $70$ &$0.1$ \\
Ref 13 & $10 ^{-8}$ & $10$ &$1.0$ \\
Ref 14,16 & $10 ^{-6}$ & $1.0$ &$0.1$ \\ 
\hline
\end{tabular}
\end{table*}

\subsubsection{Using the same equation of state in the hare and the hounds}
\begin{figure}
	\centering
		\includegraphics[width=8.5cm]{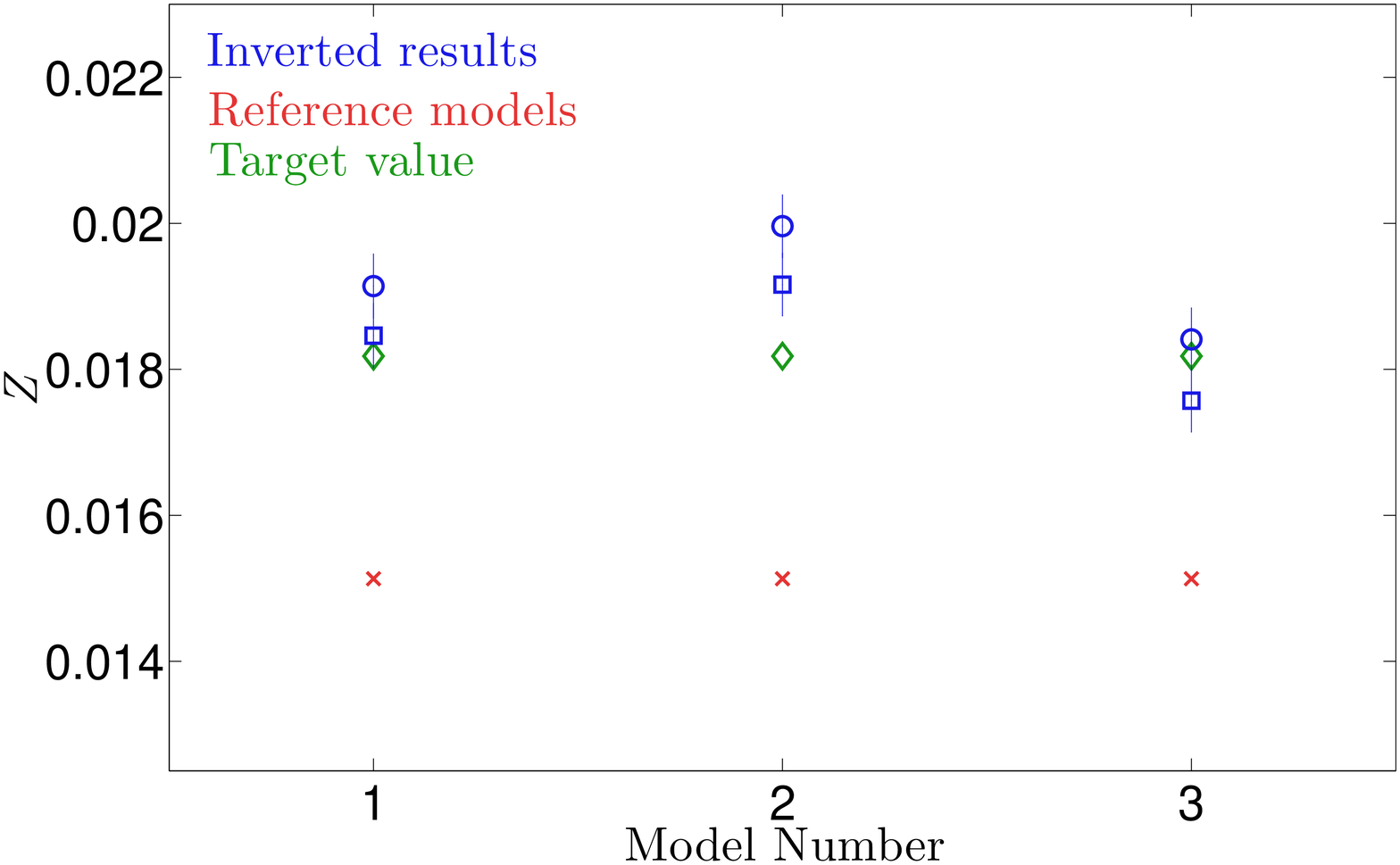}
	\caption{Inversion results for Ref $1$, $2$ and $3$ of the hare and hounds exercises. The blue squares depict inversion results for the target using the FreeEOS equation of state and the blue circles depict the results for the target using the Opal equation of state.}
		\label{figResHHEOS}
\end{figure}

\begin{figure}
	\centering
		\includegraphics[width=8.5cm]{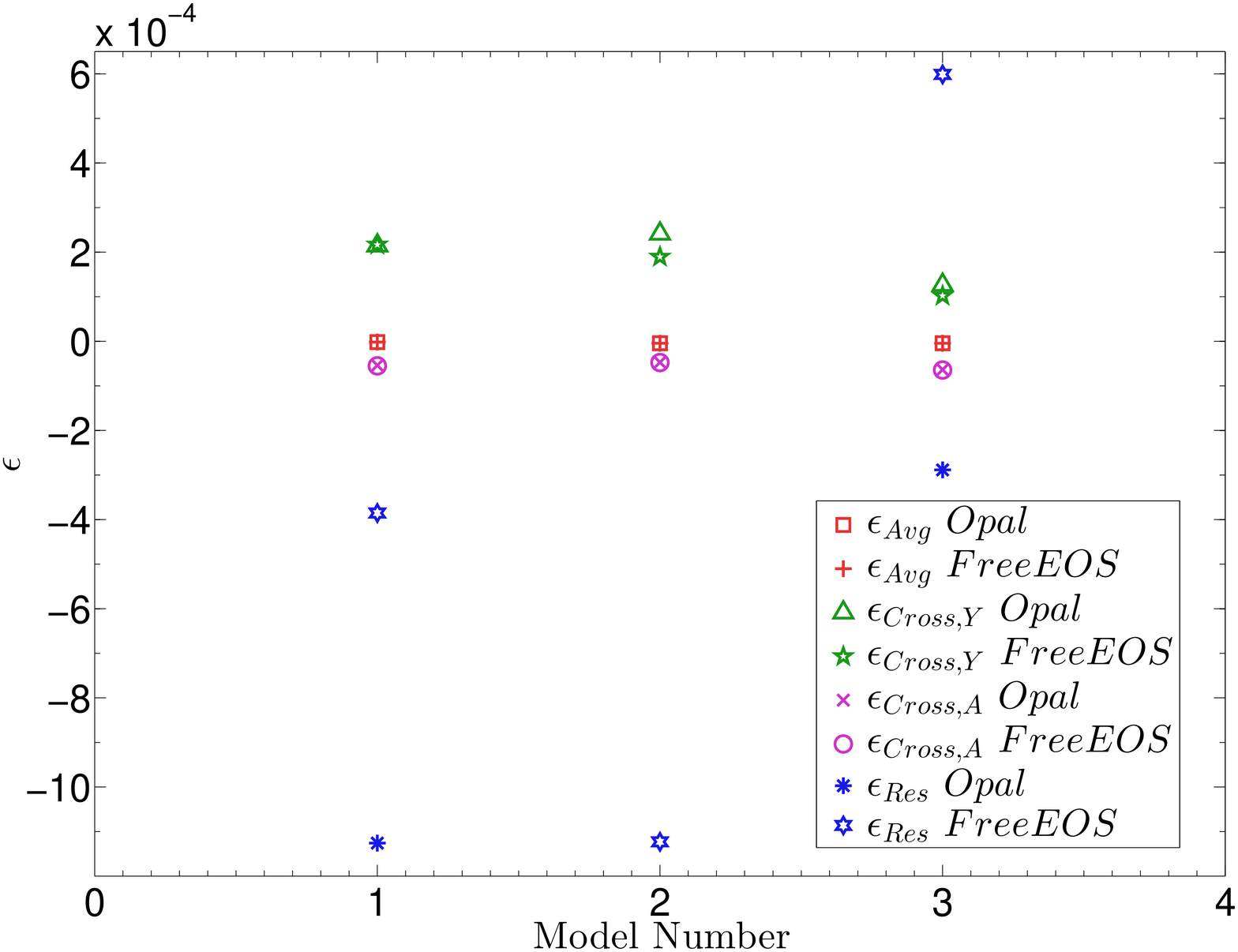}
	\caption{Error contributions, denoted $\epsilon_{Avg}$,$\epsilon_{CrossY}$,$\epsilon_{Cross,A}$,$\epsilon_{Res}$ for the hare-and-hounds exercises for both targets using Opal and FreeEOS equations of state. }
		\label{figErrorContribEOS}
\end{figure}

To quantify more reliably the effects of the equation of state, we compare metallicity inversions obtained from Ref $1$, $2$ and $3$ for a calibrated standard solar model, denoted target $2$, built using the same physical ingredients as the target $1$ of table \ref{tabHoundsProp}, with the exception of the equation of state which is the FreeEOS equation of state used in the hounds. The chemical composition of this new target is only very slightly different than the one given in table \ref{tabHoundsProp}, since the changes induced by using the FreeEOS equation of state instead of the Opal equation of state are very small. Indeed, the metallicity in the enveloppe of this new target is $Z_{CZ}=0.01820$ and its helium abundance is $Y_{CZ}=0.2386$. The inversions are carried out using the optimal set of trade-off parameters given in table \ref{tabParamsInvHH}. The inversion results are illustrated in figure \ref{figResHHEOS}, where we used circles as the notation for the inversions on the target using the Opal equation of state and squares for the inversions on the target using the FreeEOS equation of state. Due to the closeness of the value $Z_{CZ}$, for both targets, we plotted only one metallicity value to ease the readibility of figure \ref{figResHHEOS}. Similarly, the error contributions are illustrated in figure \ref{figErrorContribEOS}, where we again differentiated the two targets by using different symbols. The kernels fits are also given in table \ref{tabResInvHHEOS}.

From figure \ref{figResHHEOS}, we can see that using the same equation of state improves the inversion results for $2$ of the $3$ models. Analysing the results of table \ref{tabResInvHHEOS}, we can see that the kernel fits have not changed when compared to the values found for the same models in table \ref{tabResInvHH}. However, from figure \ref{figErrorContribEOS}, it clearly appears that the variation of the results is due to a variation of the residual error for all models. This is expected since it is this error contribution that should be affected by the change in the equation of state. For Refs $1$ and $2$, $\epsilon_{Res}$ is clearly reduced by using the same equation of state in the target and reference model. Yet, the degradation of the results for Ref $3$ illustrate that the good quality of the results obtained for the target built with the Opal equation of state might sometimes be fortuitous.

\begin{table*}
\caption{Inversion results for the hare-and-hounds exercises for the target built with the FreeEOS equation of state.}
\label{tabResInvHHEOS}
  \centering
\begin{tabular}{r | c | c | c | c |c }
\hline
& $Z_{Ref}$  & $Z_{Inv}$ & $\vert \vert K_{Avg}-\mathcal{T}_{\mathcal{Z}} \vert \vert^{2}$ & $\vert \vert K_{Cross,Y} \vert \vert^{2}$ & $\vert \vert K_{Cross,A} \vert \vert^{2}$ \\ \hline
Ref 1 & $0.01513$ & $0.01846\pm 4.35 \times 10^{-4}$ & $0.02109$ & $0.07683$ & $5.000\times 10^{-3}$\\  
Ref 2 & $0.01513$&  $0.01916\pm 4.26 \times 10^{-4}$ & $0.02198$& $0.07680$& $4.543\times 10^{-3}$\\  
Ref 3 & $0.01513$& $0.01757\pm 4.25 \times 10^{-4}$ & $0.02169$& $0.07602$ & $4.549 \times 10^{-3}$ \\
\hline
\end{tabular}
\end{table*}

\subsubsection{Changing the difference in $Z$ between the hares and the hound}
The next step in analysing the behaviour of the inversion is to see how it would behave if smaller metallicity differences were present. Indeed, in the previous test cases, we specifically built our target and reference models such that they showed clear differences in metallicity. Assessing the behaviour of the inversion if the metallicity is already close is crucial since the real solar case is slightly more complicated. Indeed, metallicity determinations have ranged from $0.0201$ in \citet{AG} to $0.0122$ in \citet{AspG}, with intermediate values found in \citet{GN93}, \citet{GreSauv} or \citet{Caffau}, and one has to be able to assess the stability of the inversion for various differences in metallicity between reference and target. This has already been done to some extent with the four first reference models, which had a higher metallicity of $0.01513$ in the convective zone.

\begin{table*}
\caption{Inversion results for the supplementary hare-and-hounds exercises between reference models.}
\label{tabInvHHRef}
  \centering
\begin{tabular}{r | c | c | c | c | c | c}
\hline 
& $Z_{Ref}$  & $Z_{Inv}$ & $Z_{Tar}$& $\vert \vert K_{Avg}-\mathcal{T}_{\mathcal{Z}} \vert \vert^{2}$ & $\vert \vert K_{Cross,Y} \vert \vert^{2}$ & $\vert \vert K_{Cross,A} \vert \vert^{2}$ \\ \hline
Ref 1 to Ref 8& $0.01513$ & $0.01163\pm 2.154 \times 10^{-5}$ & $0.01381$& $0.02214$ & $0.07680$ & $1.832$\\  
Ref 4 to Ref 13 & $0.01513$&  $0.01264\pm 2.229 \times 10^{-5}$ & $0.01373$& $0.03248$& $0.07518$& $1.765$\\  
\hline
\end{tabular}
\end{table*}

\begin{table*}
\caption{Error contributions for the supplementary hare-and-hounds exercises between reference models.}
\label{tabErrConRef}
  \centering
\begin{tabular}{r | c | c | c | c | c | c}
\hline
& $\epsilon_{Avg}$  & $\epsilon_{Cross,Y}$ & $\epsilon_{Cross,A}$& $\epsilon_{Res}$  \\ \hline
Ref 1 to Ref 8& $-3.785 \times 10^{-7}$ & $2.594 \times 10^{-4}$ & $-6.667 \times 10^{-4}$ & $2.598 \times 10^{-3}$ \\  
Ref 4 to Ref 13 & $-4.4674 \times 10^{-6}$&  $-4.301 \times 10^{-5}$ & $-5.289 \times 10^{-5}$& $1.197 \times 10^{-3}$\\  
\hline
\end{tabular}
\end{table*}

We now supplement these test cases by carrying out metallicity inversions using Ref $8$ and $13$ as our new targets. We chose Ref $1$ and Ref $4$ for respective hounds of Ref $8$ and Ref $13$. The results are given in table \ref{tabInvHHRef}. They show that smaller changes in metallicity can be seen but the inversion starts to be less reliable. This is expected since the value of the residual error, illustrated in table  \ref{tabErrConRef}, which can go as high as $3 \times 10^{-3}$. Moreover, as $\delta Z$ decreases, the importance of the contributions of the other integrals, associated with $\delta Y$ and $\delta A$ is increased, and the information on the metallicity is consequently more difficult to extract. However, these exercises also show that in such cases, the parameters leading to an accurate inversion result are slightly different than what is found before. Indeed, the parameters were then $\beta=10$, $\beta_{2}=10^{-2}$ and $\theta=10^{-3}$, with the inversion being rather stable around those values, but sometimes showing erratic behaviour since the contribution from helium or the convective parameter are much higher compared to the contribution of the metallicity in the frequency differences. Thanks to these results, we are now able to assess the quality of the inversion of the real solar metallicity.

\section{Solar inversions} \label{sec:suninv}
\subsection{Inverted results and trade-off analysis} \label{sec:sunRes}
Using the parameter sets of table \ref{tabParamsInvHH} and the analysis performed in our hare-and-hounds exercises, we can now carry out inversions on the actual solar data. We used various models calibrated on the Sun as references for the inversion. We used the solar radius, luminosity and current value of $\left(\frac{Z}{X}\right)_{\odot}$ according to the abundance tables used for the model as constraints for the solar calibration. We started with models built using the GN$93$ and AGSS$09$ abundance tables with multiple combinations of physical ingredients, such as the CEFF or FreeEOS equations of state and the Opal \cite{Opal} or OPLIB \citet{Colgan} opacity tables, summarised in table \ref{tabModSun}. It should also be noted that Model $5$ was built using a uniform reduction of $25\%$ of the efficiency of diffusion, resulting in a higher helium abundance and shallower base of the convective zone than Model $1$. Similarly, Model $7$ was built with an ad-hoc temperature-dependent modification of the diffusion coefficients of the heavy elements that led to significant changes in helium abundance in the convection zone. In addition to this modification, an undershoot of $0.15$ of the local pressure scale height was added to improve the agreement with the helioseismic determination of the base of the convection zone.
\begin{table*}
\caption{Properties of the solar models used for the inversion.}
\label{tabModSun}
  \centering
\begin{tabular}{r | c | c | c | c | c | c | c}
\hline 
& Model 1  & Model 2 & Model 3 & Model 4 & Model 5 & Model 6 & Model 7 \\ \hline
Mass $(M_{\odot})$ & $1.0$ & $1.0$ & $1.0$ & $1.0$ & $1.0$ & $1.0$ & $1.0$\\  
Age $(Gy)$ & $4.58$&  $4.58$ & $4.58$& $4.58$& $4.58$ &$4.58$& $4.58$\\  
Radius $(R_{\odot})$& $1.0$& $1.0$ & $1.0$& $1.0$ & $1.0$ & $1.0$& $1.0$\\
$L$ $(L_{\odot})$ & $1.0$ & $1.0$ & $1.0$ & $1.0$ & $1.0$ & $1.0$ & $1.0$\\
$Y_{CZ}$ & $0.2385$ & $0.2475$ & $0.2445$ & $0.2416$ &$0.2432$&$0.2292$&$0.2435$\\
$Z_{CZ}$ & $0.01820$ & $0.01803$ &$0.01809$ & $0.01813$ & $0.01805$ &$0.01370$&$0.01344$\\
$\left(\frac{r}{R}\right)_{CZ}$ & $0.708$ & $0.708$ &$0.711$ & $0.712$ &$0.711$ &$0.718$& $0.712$\\
Opacity &OPLIB& OPLIB$\times 1.05$& OPLIB & OPAL &OPLIB &OPLIB & OPLIB \\
EOS & FreeEOS & FreeEOS & FreeEOS & CEFF & FreeEOS & FreeEOS & FreeEOS \\
Abundances & GN$93$ & GN$93$ & GN$93$ & GN$93$ & GN$93$ & AGSS$09$ & AGSS$09$ \\
\hline
\end{tabular}
\end{table*}

\begin{figure*}
	\centering
		\includegraphics[width=13.5cm]{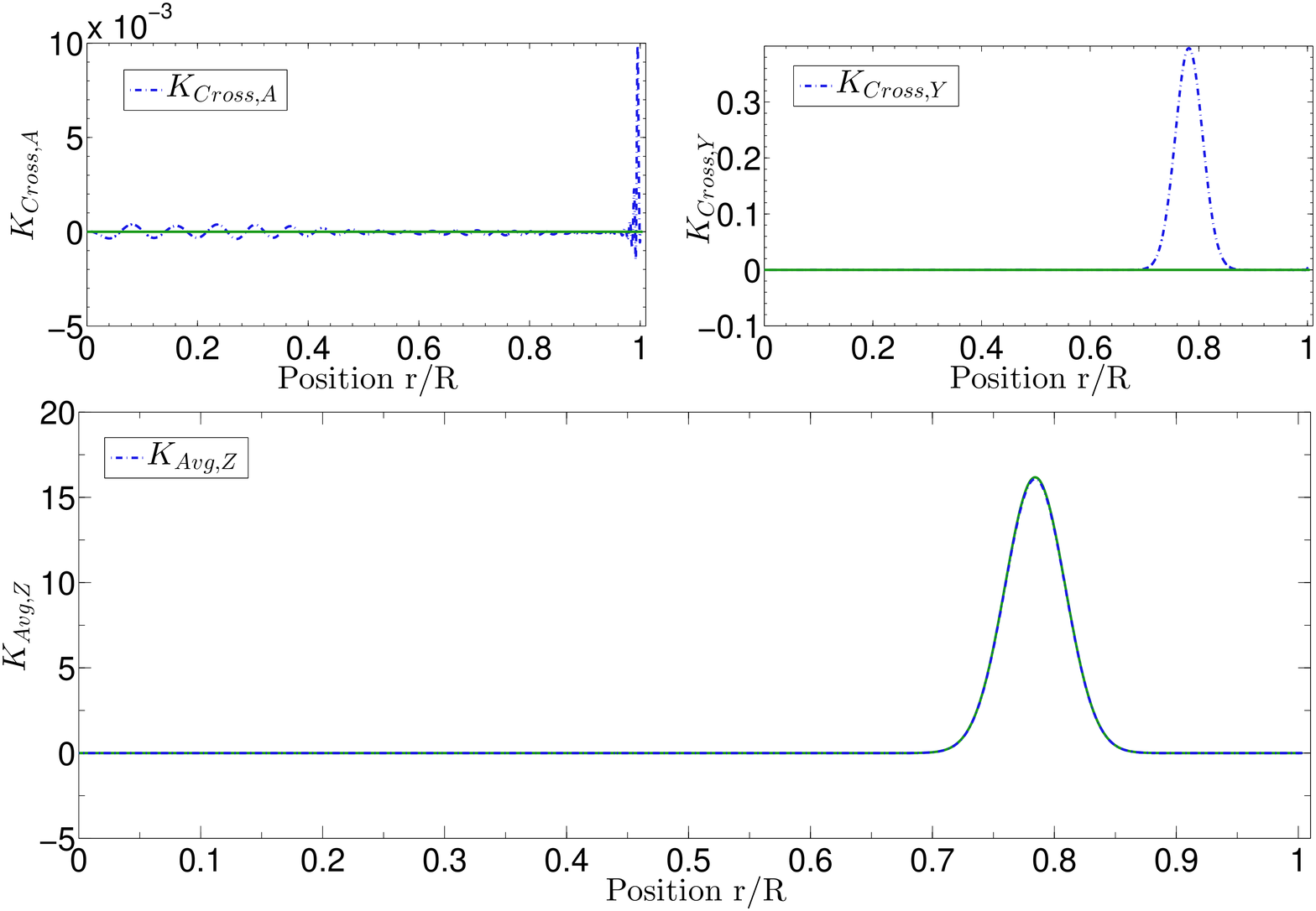}
	\caption{Same as figure \ref{figAvgCross} for a solar inversion.}
		\label{figAvgCrossSun}
\end{figure*}

\begin{figure*}
	\centering
		\includegraphics[width=13.5cm]{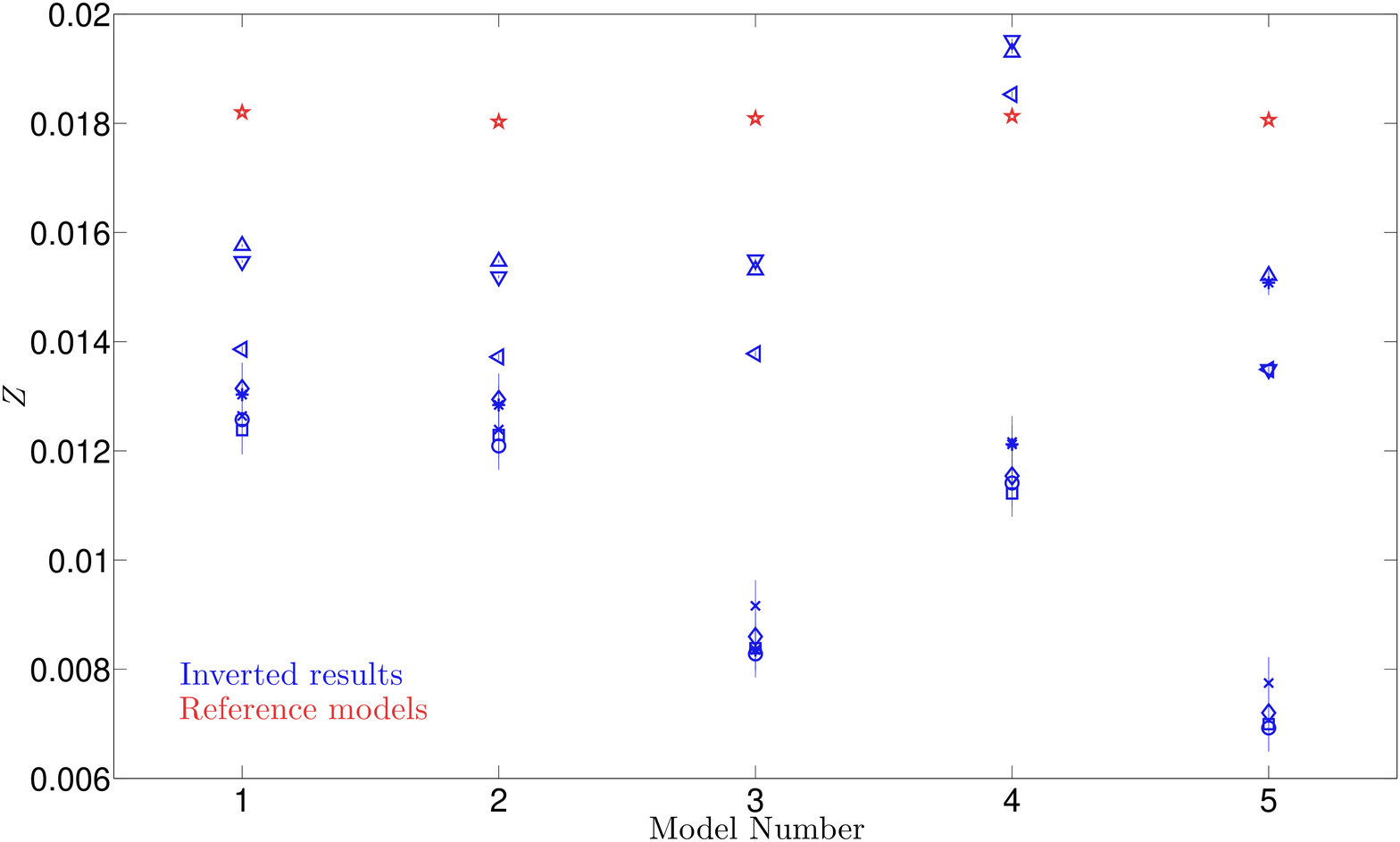}
	\caption{Inversion results for various solar models. The reference metallicity values are ploted as red $\medstar$, while the inverted values are given in blue with each symbol corresponding to a trade-off parameters set. Model $6$ and $7$ are excluded from the plot (see manuscript for further comment).}
		\label{figResZSun}
\end{figure*}

The inversions were carried out for the various combinations of parameters found in the hare-and-hounds exercises and illustrated in table \ref{tabParamsInvHH}. We illustrate in Fig \ref{figAvgCrossSun} the fits of the averaging and cross-term kernels of the inversion, showing that the quality of the inversion in the solar case is comparable to that of the hare-and-hounds exercises. The inversion results are illustrated in Fig \ref{figResZSun} for various reference models identified by their number in abscissa, and where each symbol corresponds to one set of trade-off parameters. Namely, the triangles correspond to the sets associated with higher values of $\theta$ and lower values of $\beta_{2}$, whereas the other symbols are associated with values of $\theta$ between $10^{-5}$ and $10^{-8}$.

The first result is that a very large majority of inversions favour a low metallicity, as presented in \citet{Vorontsov}. Only one small region of the parameter space, for the inversion of Model $4$, built with the CEFF equation of state, gives a high metallicity as a solution. However, this solution is not very trustworthy since it is associated with a high cross-term kernel of $A$. This high cross-term value is a consequence of the parameter set used, which is that of the less stable inversions presented in Sect. \ref{sec:harehounds}. Inversions of the convective parameter have shown that Model $4$ presented large discrepancies with the Sun in its $A$ profile, thus showing the need to damp the the cross-term contribution for this model. When such a damping is realised, the solution directly changes to a low-metallicity one. This leads us to consider that these particular results, obtained with a particular parameter set are not to be trusted. Moreover, even for the other models, this set gave higher metallicities, indicating that a systematic error, stemming from the $A$ differences with the Sun, was potentially re-introduced. Based on these considerations, we consider that the inverted results obtained with this parameter set are not to be trusted.
Ultimately, one finds out that the given interval is between $\left[ 0.008, 0.0014 \right]$. While the lowest values of this inversion seem unrealistically low, one should keep in mind that errors of the order of $3.0 \times 10^{-3}$ have been observed in the hare-and-hounds exercises, explaining the interval obtained here. This spread is a consequence of the variations in the ingredients of the microphysics in the models, which lead to better or worse agreement with the acoustic structure of the Sun.

Similar tests have been performed with low metallicity AGSS$09$ models, to assess whether they would favour a high metallicity solar envelope, as in our hare-and-hounds exercises. The results are quite different. For all parameters sets used in the first sets of hare-and-hounds exercises, the inversion gives unrealistic results, ranging from negative metallicities to values of the order of $0.1$ or more, depending on the models and parameter sets. This is not surprising since one infers directly the metallicity difference, $\delta Z$, from Eq. \ref{eqInvZ}. This means that if this contribution is very small, it will be dominated by the other differences, $\delta A$ and $\delta Y$. However, in Sect. \ref{sec:harehounds}, we attempted a few inversions for smaller metallicity differences between targets and reference models. We showed that in some cases, it was possible to infer the result but that the accuracy was reduced and the cross-term could dominate the results. Using the same parameters as in our supplementary hare-and-hounds exercises, we find that the AGSS$09$ models place the solar metallicity value around $1.1 \times 10^{-2}$, which is in agreement with the previous results.

The differences in  acoustic  structure  are expected to be larger  between  the Sun and  the  low  metallicity  standard  models  than  between two numerical models of various metallicities considered in our H$\&$H exercises. Hence, it is not surprising that the inversions performed in our H$\&$H exercises are more successful than real solar inversions of low metallicity standard models. The large differences imply larger cross-term contributions, which stem mainly from two regions. The first contribution comes from the radiative zone, especially the region just below the convective envelope, where the differences in $A$ can be quite large. The second contribution is due to the thin region of the convective envelope where the non-adiabaticity of the convective transport increases and thus $A$ has values strongly different than $0$. Another weakness of low metallicity standard solar models is that they do not reproduce the helium abundance in the convective zone. For example, the value found in Model $6$ could be different from the solar value by up to $0.025$, implying that even if the helium cross-term kernel is strongly damped, it could still strongly pollute the results of the solar inversion. In contrast, the models used in the hare-and-hounds exercises were calibrated to fit the helium value within an accuracy of $0.01$. This problem was less present for Model $7$, for which the ad-hoc changes in the ingredients lead to a better agreement with helioseismic constraints and consequently, the inversion was slightly more stable. However, the sound speed, density and convective parameter profile inversions still showed larger differences than for GN$93$ models despite the changes. Therefore, we deem the reliability of the solar inversions from the AGSS$09$ models, especially that of Model $6$, which is particularly unstable, to be less robust and constraining as the inversions from the GN$93$ models. Therefore we excluded these models from Fig \ref{figResZSun}.

Despite these uncertainties, the indications obtained from these models for a restricted region of the parameter space, combined with the fact that the inversions from GN$93$ solar models favour a low solar metallicity, still lead us to advocate for a low value of the solar metallicity. However, the precision of this value is quite poor and limited by the intrinsic errors in $\Gamma_{1}$ and the uncertainties on the equation of state. Still, we can conclude that our results do not favour the GN$93$ abundance tables and any other table leading to similar solar metallicities, like the GS$98$ or AG$89$ tables. Had the solar metallicity been that of these abundance tables, a clear trend should have been seen when carrying out solar inversions from low-metallicity AGSS$09$ models.

\subsection{Consequences for solar models} \label{sec:Disc}

The results presented in section \ref{sec:sunRes} have significant consequences for solar modelling since they confirm the low metallicity values found by \citet{Vorontsov}. This result implies that the observed discrepancies found in sound speed inversions \citep[e.g.][]{SerenelliComp} are not to be corrected by increasing the metallicity back to former values, but could rather originate from inaccuracies in the physical ingredients of the models. Amongst other, the opacities constitute the physical input of standard solar models that is currently the most uncertain. Recently, experimental determinations of the iron opacity in physical conditions close to those of the base of the solar convective zone have shown discrepancies with theoretical calculations of between $30 \%$ and $400 \%$ \citep{Bailey}. The outcome of the current debate \citep[see for example][amongst others]{Nahar, BlancardComment, IglesiasIron} in the opacity community that these measurements have caused will certainly influence standard solar models and the so-called ``solar modelling problem''.

Besides the uncertainties in the opacities, the inadequacy of the low-metallicity solar models could also result from an inaccurate reproduction of the sharp transition region below the convection zone. In current standard models, this region is not at all modelled, although it is supposed to be the seat of multiple physical processes \citep[see][for a thorough review]{Hughes2007} that would affect the temperature and chemical composition gradients, and thus the sound speed profile of solar models.

In addition to these main contributors, further refinements of physical ingredients such as the equation of state \citep[see for example][for recent studies]{BaturinSAHA,Vorontsov13} or the diffusion velocities could also slightly alter the agreement of low-metallicity solar models and the Sun \citep[see for example][for a study of the effects of diffusion]{Turcotte1998}.

\section{Conclusion} \label{sec:Conc}

In this paper, we presented a new approach to determine the solar metallicity value, using direct linear kernel-based inversions. We developed in Sect. \ref{sec:method} an indicator that could allow us to restimate the value of the solar metallicity, provided that the seismic information was sufficient. We showed that the accuracy of the method was limited by intrinsic errors in $\Gamma_{1}$ and differences in the equation of state.

In Sect. \ref{sec:harehounds}, we carried out extensive tests of the inversion technique, using various physical ingredients and reference models. These tests showed that while the inversion could distinguish between various abundance tables such as those of AGSS$09$ and higher metallicity abundances, like those of GN$93$ or GS$98$, it could not be used to determine very accurately the value of the solar metallicity, due to intrinsic uncertainties of the method and models. 

In Sect. \ref{sec:suninv}, we applied our method on solar data, using the inversion parameters calibrated in the hare-and-hounds exercises carried out previously. We concluded that our method favours a low metallicity value, as shown in \citet{Vorontsov}. However, further refinements to the technique are necessary to improve the precision beyond that achieved in this study.

Firstly, developing a better treatment of the cross-term contribution from the convective parameter, $A$, taking into account its $0$ value in the convective zone, could help with the analysis of the trade-off problem. Secondly, using other equations of state could help with understanding the uncertainties this ingredient induces on the inversion results. Thirdly, the use of a seismic solar model based on inversions of the $A$ and $\Gamma_{1}$ profiles as a reference, rather than calibrated solar models, could reduce the impact of the intrinsic $\Gamma_{1}$ errors, thus reducing the main contributor to the uncertainties of the inversion and provide more accurate and precise results for this inversion.  

\section*{Acknowledgements}

G.B. is supported by the FNRS (``Fonds National de la Recherche Scientifique") through a FRIA (``Fonds pour la Formation \`a la Recherche dans l'Industrie et l'Agriculture") doctoral fellowship. This article made use of an adapted version of InversionKit, a software developed in the context of the HELAS and SPACEINN networks, funded by the European Commissions's Sixth and Seventh Framework Programmes.




\bibliographystyle{mnras}
\bibliography{biblioarticleZ} 

\bsp	
\label{lastpage}
\end{document}